\documentstyle[titlepage,fleqn,12pt]{article}
\input epsf
\begin{document}

\begin{center}
{\large A possible scenario for volumetric display through nanoparticle suspensions 
} \\ 
\vspace{0.5cm}

\vspace{0.5cm}
Enrique Canessa \\
{\it The Abdus Salam International Centre for Theoretical Physics, Trieste, Italy}
\end{center}

\vspace{2cm}
{\baselineskip=22pt
\begin{center}
{\bf Abstract}
\end{center}

We discuss on the potential of suspensions of gold nanoparticles with 
variable refractive index for the possible physical realization of 
{\it in-relief} virtual dynamic display of plane images.
A reasoning approach for a vision system to display in real-time 
volumetric moving images is proposed based on well-known properties 
of optical media, namely the anomalous dispersion of light on certain 
transparent media and the virtual image formed by a refracting transparent
surface.  The system relies on creating mechanisms
to modify the refractive index of {\it in-relief} virtual
dynamical display (iVDD) bulbs that ideally would contain a
suspension of gold nanoparticles each and that might be ordered
in an array filling up a whole screen.

}

\baselineskip=22pt
\parskip=4pt

\pagebreak

In this work we present an approach for a vision system 
based on suspensions of gold nanoparticles with variable refractive index 
that could allow to depict in real-time and {\it in- relief}
two-dimensional 2D ({\it i.e.}, planar or flat) virtual images such as light rays 
that do not
converge to the image point but appear to emanate from that point
that could be in movement.

To perceive, recognize and gauge with our brain the {\it depth} of a
3D picture, 
nature provides us with two images of the same scene (the so-called
{\it stereopsis} process \cite{Bra83,Ser96,Fal86,Bor75}).
The amount of depth perceived also depends on the
binocular disparity between the two eyes views.  The wider the separation,
the greater the apparent depth is.  By moving our head and looking around
objects, we change our view and judge depths of objects at different distances
from our eyes (the {\it parallax} phenomenon).  
The brain also uses the so-called {\it accommodation} of blurred images
to gauge depth from the muscular contraction of the {\it lens} surface
of the eyes.

Besides the input signals from the two eyes, the brain also uses other cues
in the determination of depth.  These include the cues that artists rely on
to convey a feeling of depth in 2D paintings.
Artists convey the roundedness of an object by varying color
or indicating the different light intensities on different surfaces
({\it e.g.}, shadows).  Flat, static pictures can also take on some depth
by variations in the sharpness of distant objects ({\it e.g.},
perspective) or by the interposition ({\it i.e.}, occlusion) of objects.

Separate images to each eye can also present a binocular disparity and hence
a 3D view to the observer ({\it e.g.}, as seen by binoculars).  To this
end, the stereoscope presents two different pictures to each eye whose
viewing appears in depth due to their differences.   This old-technique has been
used, {\it e.g.}, in Astronomy to notice changes in the stars positions.
Other attempts for achieving static 3D displays are in-relief posters and
maps, multiple-layer postal cards, etching monographs, embossed image
art, sculptures, holograms (to record and store 3D visual information
to be re-displayed under proper illumination), {\it etc}.

The shape-from-shading (also known as {\it photoclinometry}) is a method
used to view planar images in 3D by mathematically relating image
intensities and brightness \cite{Hor77}.  Other mathematical techniques
make it also possible to reconstruct from 2D projections and get a full 3D
picture \cite{Gor75}.

Alternative techniques to visually fathom the depths of the world in
movement around us have been also explored.  These include the
3D-like Cinevision with added external effects and objects ({\it e.g.},
perfume, dummies and quakes superimposed), the virtual reality
(with simulated virtual touch) and 3D computer vision \cite{Bre98}
(based on perspective and light intensity changes and motion sensors).
Related work on the 3D display using non-linear optics
can be found in \cite{Dow94}.

Motivated by all this firestorm of ongoing activity on the search of
techniques for giving the impression of reality (depth) and hopes
for subsequent realistic displays, we focus next in the possible vision 
system based on suspensions of gold nanoparticles with variable 
refractive index.  
To achieve an enhanced depth by the {\it in-relief} virtual
dynamical display (iVDD) of plane images, it is necessary
to consider first two physics phenomena as seen in nature and described
by geometrical optics (apparent depth) and nonlinear optics (anomalous
refraction index.

Images of objects inmersed in a medium with refraction index $n_{1}$ are
formed by the refraction of rays at a, say, spherical surface with radius
$R$ of a transparent material
with a different $n_{2}$ \cite{Ser96}.
By applying Snell's law to the refracted rays
and after a simple geometrical construction, it can be shown that
$\frac{n_{1}}{p} + \frac{n_{2}}{q} = \frac{n_{2}-n_{1}}{R}$,
where $p$ is object distance and $q$ is image distance from the
spherical surface.  For a flat surface, one approximates $R$ to infinite
hence for small angles
\begin{equation}\label{eq:num1}
q \approx -(n_{2}/n_{1})p \;\;\; .
\end{equation}
When the sign of $q$ is opposite to that of $p$, the location of
the virtual image is on the same side of the surface as the object.

It is well-kwown that the refractive index (or index of refraction),
characteristic of a substance, varies with light.  If $n$ changes we then
deduce from Eq.(\ref{eq:num1}) that the position of a virtual image might
move as a function of the frequency of radiation $\lambda$.  
The nonlinearity of the refractive index versus $\lambda$ in certain optical
media as well as
the connection between {\it dispersion} and {\it absorption} has been known
for many years \cite{Par88}. (For details of the treatment
of dispersion according to quantum theory and classical Lorentz electron
theory see, {\it e.g.}, \cite{Bor75,Akh97,Bal69,Zer73,Ber71}).

{\it Absorption} involves the exitation by radiation of an electron from
its lowest energy state to a higher energy state by the incident light wave
(quantum confinement).  Usually not all of the absorbed
energy is re-radiated, remaining as heat in the absorbing material.
In quantum theory, absorption is associated with transitions between
quantized energy levels.  Whereas according to classical theory, this
phenomenon is identified with the steady oscillation of a charge in an
orbit reacting to radiation.  Each electron oscillator has a resonance
frequency and, near resonance, the damping oscillator produces the
broadening of the amplitude response.

The refraction or bending of a ray of light that passes through a
medium is the result of the diminished transmission rate of the wavefronts
in the medium.  The velocity of light propagation in the medium varies
considerably with wavelength $\omega$ hence the refractive index
$n$ varies with the wavelength of the incident light.
When $\omega$ approaches
the wavelengths of strong absorption bands of the medium, $n$ undergoes
sharp changes around resonant frequencies $\lambda_{R}$.  

This anomalous
behavior of $n$ was first observed in electrically excited rarified gases
some time ago (see reviews in \cite{Kor32,Lad33}).  Since then, there
has been a continuing interest in investigating the exceptionally high
values of the refractive index near resonance lines by laser excitation
in bulk semiconductors
\cite{Gib90,Ric90} and Sodium vapor \cite{Kug83} to mention a few.
Experiments in air and diatomic gases show that the resonant wavelengths
lie in the ultraviolet ({\it i.e.}, shorter than the visible wavelengths).
The characteristic damping times for the resonances are usually compared
with the incident light intensity pulses used.

Variations in density $\rho$ (due to, {\it e.g.}, compressibility)
\cite{Ber71}, in small temperature gradients \cite{Fie85} and in angle
of light incidence \cite{Lyn95} also produce changes in
optical refractivity.  Though such changes are not sufficient to produce
appreciable percentage of excited states (and produce the so-called
{\it `negative dispersion'} effect).

Within classical electromagnetic theory, $n$ is represented as a complex,
frequency dependent quantity \cite{Akh97,Ber71}.  It can be shown that
$n(\omega )=Re \sqrt{\epsilon (\omega )}$, where $\epsilon$ is the
dielectric permittivity of the medium, and the absorption coefficient is
given by $\delta(\omega )=-2(\omega /c) Im \sqrt{\epsilon (\omega )}$.

There is also a close relation between
dispersion and absorption.  For a Lorentzian oscillator, the extreme
values of $n$ occur at a frequency at which $\delta$ has half its maximum
value and it can be approximated by \cite{Ber71}
\begin{equation}\label{eq:num2}
(n-1)_{extreme} \approx n_{o} - 1 \pm \frac{\delta_{max}}{2}\;\;\; ,
\end{equation}
where $n_{o}-1$ is the reduced refractivity due to different electron
oscillators per molecule.

It is important to mention that measurements of nonlinear refraction at room
temperature and upon ($\sim$picosecond) excitations by a laser pulse have
also been reported in porous-glass nanocomposite materials \cite{Geh97,Uch94}
and transparent dielectrics materials \cite{She90} owing to their potential
applications for improving the performance of photonic devices.
On the theoretical side, the electronic
properties of these particles with diameters of few up to ten nanometers
and embedded in a dielectric have been studied in \cite{Sta98,Per81}.

A large amount of work has also been devoted to the study of quantum size
effects in suspensions of submicron size colloidal particles \cite{Per81}.
However, it has been only in recent years that the ultrafast nonlinear
response of metal colloids to light has been measured (whose diameters
are much smaller than the optical wavelength) \cite{Blo90,Ols91,Meh97}.
These fascinating nanoparticle systems, which are neither quite microscopic
nor quite macroscopic \cite{Per81}, are of our interest for the iVDD
implementation and the possible 3D display of visible plane surfaces.

Experimental results using a single beam
technique \cite{Shi90}, show that the non-linearity is {\it ultra-fast} with
response time shorter than, {\it e.g.}, the $28\; ps$ pulse duration of
the laser used.  The absorption spectrum of gold nanoparticles having
mean diameter of $\approx 20\; nm$ displays a maximum at a frequency of
$\approx 525\; nm$.  Particles larger in size have a red-shifted
plasmon maximum \cite{Ahm96}.
Copper nanoparticles in 2-ethoxyethanol and water shows distinct
absorbtion peaks at 572 and 582 nm, respectively \cite{Hua97}.

The optical response of rod-shaped gold nanoparticles in solutions, with
different aspect ratios, is also of interest in the visible \cite{YYYu97}.
It has been observed that their absorption spectral features consist of
a dominant surface plasma attributed to the collective
dipole oscillation. corresponding to the longitudinal resonance.
Its characteristic frequency shifts to the red when increasing the
 anisotropic shape of
the nanorods.  As with the nanoparticles in suspension, data from human
photoreceptors in the retina shows that the wavelength of maximum absorbance
of the sensitive receptor cells known as cones lies also in the range
$534-564 \;nm$ \cite{Bow80}.

As anticipated the initial stages of image processing in nature consist
of fairly simple, usually local, transformations of the array of (refracted
light) intensity values in the image \cite{Bra83}.  Inspired by these
observations, in what follows, we introduce the iVDD system for the
image filter that might produce the sensation of depth under normal
viewing conditions without the use of external glasses or special sensors.

An input signal (such as a TV signal comming from an antenna) could be
simultaneously displayed in two receiver sets.  A radiation-sensitive detector
photodiode that converts light into an electrical signal,
placed in front of one of the TV screens, could be used for the voltage
analysis of the time-varying imagery in real time.  The photodiode could
send a voltage signal to a selective switch.  The switch could possess
two different (or a discrete set of) output states of emitted light that
could radiate the iVDD bulb in the visible portion of the electromagnetic
spectrum in which the moving images containing 2D information transmitted
by a second receiver set might cross.

Following reported experimental data,
if frequencies at or near resonance $\lambda_{R}$ can be used to radiate
the iVDD bulb, sufficient excitation of energy states would produce a
nonlinear $n$ of the suspension.  Since the refractive index of the iVDD
bulb (filled with the nanoparticles) would change abruptly then one can deduce
from Eq.(\ref{eq:num1}) that the position of a virtual image should move
(back or forward) as a function of the frequency (greater or smaller
than $\lambda_{R}$).  The incident wavelengths should be near (and
not equal to) the wavelength at which the nano gold colloidal suspension
has a {\it maximum} absorption.  So little should be
transmitted or reflected of the incident light in a similar way to
fluorescent-tube discharge lamps where the incident UV light excites
a fluorescent coating on the internal surface of a bulb.

A coherent laser light could be activated by the voltage signals coming
from the photodiode.  The required high-frequency radiation could be
produced by, {\it e.g.}, a Nd:YAG laser ($\lambda \approx 530 \;nm$)
providing $30 \;ps$ pulses at a repetition rate of $10 \;Hz$
\cite{Blo90}. (For visible light, one requires frequencies of the
order of $3 \;10^{6} GHz$: since
$\lambda =c/\omega \approx 3\; 10^{8}/100\; 10^{-9}$ cycles per
seconds.  The use of Incandescent and fluorescent lights might also
emit light in this frequency range.  Besides, acoustic-optic modulation
might alter the refractive index as observed in recent experiments in
optical fibers \cite{Kat95,Tow96}.

Because the refractivity should vary sharply over the profile of a spectral
line due to the smallness of the particles inside the iVDD bulb, the
propagation of the two wavelengths of light through the iVDD bulb will
cause the position of a virtual image formed in the iVDD bulb to
move as a function of the frequency of radiation according to refraction
phenomena.  This effect would allow to simulate spatial
information out of the planar images.  By using reported experimental
data for the absorbance spectra of gold particles in suspension, we carry
out a crude estimation for the extreme value of $n$ given in
Eq.(\ref{eq:num2}) and found that the virtual image would have
(maximum, minimum) depths of about $\pm 15\%$ from the receiver screen distance.

Thus, our simple system involves the generation of light excitation
as emitted, {\it e.g.} from a TV screen.  This radiation could be used to alter the refractive
index of a stable inhomogeneous nanoparticle dispersion which generates an
{\it in-relief} virtual dynamical viewpoint and allows to gain depth
control by the light-frequency filtering.  Due to the fast-response time
of the non-linear dispersion, motion changes should appear as in
TV and cinema motion which consists of a permanent stream of frame pictures.
The electron beam fired out of a gun inside a TV set hits
a phosphor-coated screen to draw a picture at a rate of 30 times every
second. 

In principle to transfer the full screen information one could think
of an array of iVDD bulbs filling up the screen area. Each screen pixel
to be seen {\it in-relief} should be in correspondence (and be representable)
with the 2D full screen.  The viewer would see a virtual image
which might appear as a relief image of the original full-screen object.

The present system needs to be compatible with TV broadcast
standards and standard home TV receivers, including the new
digital TV broadcasting which allows to improve the overall quality of
the picture (overcoming the problems of ghosting and compressing the signal)
\cite{Mor97,Rei98}.
If we consider a (black and white) rectangular computer screen (roughly
25{\it cm} diagonal) having an array of 640 by 480 pixels, then one would
need about 307,200 iVDD bulbs to cover out all the screen.

Of course, the present ideas are only speculation
as there are many technical problems to consider in practice.  A few of
these are discussed next for completeness.

To the best of the Author's knowledge there does not exist a
similar proposal for a possible iVDD system described in the
literature (combining both nonlinear optics phenomena and simple laws
of geometrics optics as used here), or a similar product
available in the market, or a similar claim in the USA patents
database \cite{Can98}. 

Although certain significant properties of optical media have
long been studied by specialists, the techniques and instruments which
could make them observable ({\it e.g.}, precision and sensitivity of
measurements) have been only available in recent years.  These include small,
high-tech lasers (with low power requirements, low phase noise and superior
frequency stability).
The high light frequencies needed for the iVDD system are not yet achieved
by commercial {\it Voltage Controlled Oscillators} (VCOs) which are
widely used for wireless communications. These are available in radio
frequencies spanning 100 MHz to 3 GHz with supply voltages ranging from 3
to 12 Volts and operative over a wide range of temperatures.
Tiny Crystal Clock Oscillators allowing fast delivery of many frequencies
are only available in this range.  All this technology is developing.

Put more precisely, it is only now possible that these thoughts
for a 3D virtual dynamic image could become feasible.
To have a suitable experimental test platform, the
system relies on creating reliable mechanisms to modify the
iVDD bulbs's refractive index displaying the recent observed non-linear
optics properties of small metal particles in suspensions which could be
ordered in an array to fill up a whole screen.

In theory the present vision system for displaying in real-time
and {\it in-relief} 2D moving images should overcome the following
problems:
One should search for the best transparent material for the iVDD bulb
container of the nanoparticles in suspensions (made of lens; glass bounded
by smooth "curved" surfaces).
One should identify which thin optical elements to use to reduce
aberration (if any will appear).
It would be necessary to control the size distribution as well as
the stability of the gold nanoparticles in suspension against agglomeration
(percolation) and sedimentation.  It is well-kwon that grafted polymers
represent an interesting alternative means to obtain a solid matrix
and control interparticle separation.  Gold-graft copolymer composites
could then be used to characterize the distribution and spatial
arrangement of the nanoparticles.
One should avoid using {\it photorefractive}
materials -the subclass of nonlinear optics materials in which light emerging
from such substances does not necessary have its frequency proportional to the
incoming light.  It is necessary to investigate the characteristic frequencies
(leading to transition on the visible) of the optical transparent media
to be used.
One should identify the optimal visual angle and the possible
range of vision.
The iVDD should be isolated from vibration.
Small vibrational disturbances on the iVDD bulbs might cause the
outgoing laser transmission beam to deviate from the desired pointing angle.
One should quantify the iVDD (fast) response times to control and avoid
superimposition of images.
One should avoid to have images magnified upright (M={\it 
image height}/{\it object height}).
One should avoid double refraction (through birefringent substrates) in
optically anisotropic media.
One should identify if there is any absorbed energy not re-radiated,
remaining as heat in the absorbing material.  

To conclude we would like also to mention that laser-induced
holographics TV, ({\it i.e.}, moving holographic videos with a wide-angle
display of 3D dynamic images) is being developed by the MIT's Media
Laboratory whose usage is expected by the year 2020 \cite{Bai97}.
Our iVDD system could differ from Holographics TV by the following.
In principle, the iVDD system could display "mezzo" relievo images only and
not a full volumetric, 3D display as in holographics TV.
It could display in-relievo images over a reduced angle-range vision
-holographic video display is capable of rendering full-color
25x25x25mm images with a $15^{o}$ view-zone at rates over 20 frames per
second.
The iVDD system could contain less information than holograms
and could also require a Laser source.  However, it might require
 simpler manipulation as it could be easy to attach to present
TV or Computer screens.

In order to get visual effects that might look like the real things one
could try the iVDD.
It could be a reasonable, cost-effective alternative to holographic TV because
the costs of large optics necessary for holographics TV could be reduced to
acceptable levels.
Our iVDD system could also be compatible with display media other than TV,
{\it e.g.}, {\it in-relief} Computer and, probably, Cinema screens.
Many practical applications can be anticipated including the entertainment
and cultural business. The iVDD system could help to medical imaging in real
time, robotics, design, {\it etc} as well as to preserve realistic vision
of extincted species and/or rare objects of which only photographic records
exit.

It has been said that {\it virtual images cannot be displayed
on a screen} (see, e.g., Ref.\cite{Ser96}).  However,
a reasoning approach indicates that there could be a way for achieving this optical
phenomenon on a large screen made of an array of tiny 
Virtual Dynamical Display (iVDD) bulbs containing gold nanoparticle suspensions.
One could also think of the two different input images to be displayed in
two separated planes and that the resulting image disparity would be sufficient
to create a steroscopic view to cross the iVDD bulbs.

\end{document}